\documentclass[useAMS,usenatbib]{mn2e}
\input epsf
\usepackage{times}

\newcommand{\xmm} {{\sl XMM-Newton}}

\newcommand{\rosat} {{\sl ROSAT}}
\newcommand{\cl} {Abell\,194}

\title[Abell 194]{3C\,40  in Abell\,194 :\\ Can tail radio galaxies exist in a quiescent cluster?}

\author[Sakelliou, Hardcastle, \& Jetha]{Irini Sakelliou$^{1}$, M.J. Hardcastle$^{2}$, and
     N.N. Jetha$^{3,4}$ \\
$^{1}$Max-Planck-Institute f\"{u}r
Astronomie, K\"{o}nigstuhl 17, D-69117, Heidelberg, Germany \\
$^{2}$School of Physics, Astronomy and Mathematics, University of
Hertfordshire, College Lane, Hatfield AL10 9AB, UK \\
$^{3}$School of Physics and Astronomy, University of Birmingham,
    Edgbaston, Birmingham B15 2TT \\
$^{4}$CEA/Saclay, Service d'Astrophysique,
                L'Orme des Merisiers, B\^{a}t. 709,
                91191 Gif-sur-Yvette Cedex, France 
}

\begin{document}

\pagerange{\pageref{firstpage}--\pageref{lastpage}} \pubyear{2005}
\maketitle

\label{firstpage}

\begin{abstract}

The nearby cluster \cl\ hosts two luminous, distorted radio
galaxies. Both reside within the cluster's core region, being
separated in projection by only 100 kpc. It is often suggested that
tailed radio galaxies such as these reside in clusters that are under
formation and are accreting new material from their outskirts. In this
paper we study the intriguing appearance of Abell 194, and test
whether the cluster and radio source dynamics are consistent with the
cluster formation/merger model. We analyse data from the \xmm\
satellite and previously unpublished observations with the Very Large
Array (VLA), as well as presenting new data from the Giant Metre-Wave
Radio Telescope (GMRT). 

The shape of the jets, and
the lack of significant stripping of the galaxies' interstellar media, indicate that the radio galaxies are not moving at the large velocities
they would have had if they were falling into the cluster from its
outskirts; galaxy velocities of $\le300\,{\rm km \ s^{-1}}$ are adequate instead. A plausible scenario that could explain the observations is 
that the dynamics of the cluster centre are relatively
quiescent, with the dominant system of massive galaxies being bound
and orbiting the cluster centre of mass.  For plausible jet/plume
speeds and densities and the galaxy dynamics implied by this picture
of the cluster, we show that the observed jet structures can be
explained without invoking a major cluster merger event.

\end{abstract}


\section{Introduction}

Radio galaxies in clusters normally have distorted morphologies, due
to the interactions of the jets with the intracluster medium (ICM). In
particular, the jets or plumes of two classes of radio galaxies that
are found in groups or clusters of galaxies [the wide-angle tail
(WAT), and the Narrow-angle tail (NAT) radio galaxies] are bent
symmetrically into wide or narrow C-shapes. It is now believed that
this particular shape is a result of the ram pressure exerted onto the
jets and plumes as the host galaxy moves relative to the ICM. In this
case, the bending direction of the tails should indicate the direction
in which the host galaxy is moving. This information has been used in
the past to study the motion of galaxies in clusters. For example,
O'Dea et al.  (1987), by measuring the orientation of the tails in
NATs, found that the radio galaxies were associated with cluster
galaxies in isotropic orbits, while Sakelliou \& Merrifield (2000)
found that the WATs in their sample were mainly on radial orbits
heading towards or away from the cluster centre.

Additionally, it has been suggested that both types of tailed sources
are hosted by merging clusters (e.g., Pinkney, Burns \& Hill 1994).
This was believed to be required in order that the galaxy velocities
relative to the ICM should be large enough to produce the observed
bending of the radio tails. 
However, this argument depends on the composition of the jet or plume.
If the material in the plumes of WATs is light with respect to the
surrounding medium (e.g, Hardcastle, Sakelliou \& Worrall 2005), the
bending of the tails can be produced even with low galaxy velocities. 

One of the rare clusters hosting more than one radio galaxy close to
the cluster centre is Abell\,194\footnote{\cl\ is at
z=0.018 (NED). Throughout this paper we use $H_0=71~{\rm km \
s^{-1} \ Mpc^{-1}}$, $\Omega_{\rm M}$=0.3, and $\Omega_{\rm
\Lambda}$=0.7, giving a scale of 0.361 kpc arcsec$^{-1}$}.   
In this paper we use X-ray and radio observations of its core region
to explore the dynamics of the cluster and its radio galaxies. In
Sect.\ 2 we discuss the properties of the cluster, based mainly on
X-ray data obtained with the \xmm\ satellite. We present the radio
data and discuss the characteristics of the radio galaxies in
Sect.\ 3. Finally, in Sect.\ 4, we discuss a possible dynamical
scenario that can give a good description of the data.

\section{The Abell\,194 cluster}

\begin{figure}
\begin{center} 
\leavevmode 
\epsfxsize 1.0\hsize
\epsffile{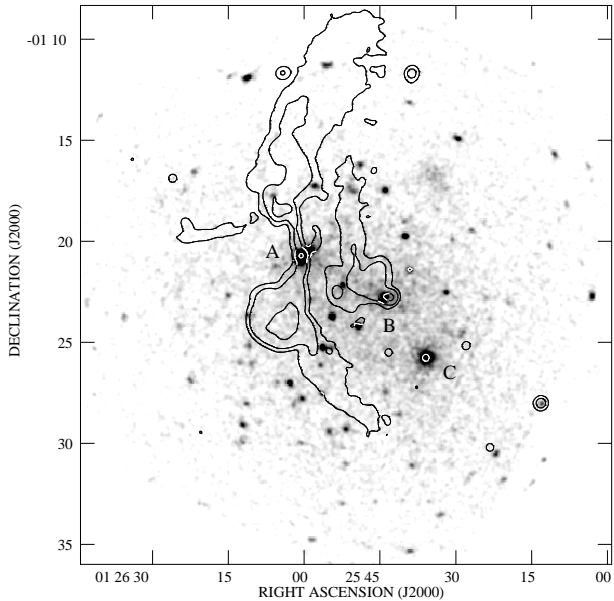}
\caption{The \xmm\ image of the cluster overlaid by the radio contours. The bright sources A, B, C
  are marked (see text). } 
\label{xray}
\end{center} 
\end{figure}

Being at a low redshift, \cl\ has been the subject of a number of
investigations and the target of observations in wavelengths ranging
from the optical to the X-rays. All past studies have found that it is
a poor and cold cluster of richness R=0. One of its intriguing
properties is that it appears as a `linear cluster'. Nikogossyan et
al. (1999) used the then available optical galaxy and \rosat\ data to
derive and compare the structural parameters of the galaxy and ICM
distributions. They found that both distributions have similar
ellipticities and position angles. Such a discovery could naturally be
interpreted as showing that \cl\ has been recently assembled by the
accretion of smaller mass units along filaments, making it a recent merger
remnant. However, searches for signatures of sub-structure related
to a recent merging activity have proved unsuccessful. To our
knowledge, both Lazzati et al. (1998) and Nikogossyan et al. (1999)
applied a wavelet analysis to the \rosat\ images, and detected X-ray
emission from some cluster galaxies and some background groups or
clusters of galaxies that appear at projected radii $>$20\,arcmin from
the core region of \cl. But localized, enhanced X-ray emission, which
would indicate the presence of dense clouds of gas within the cluster,
was not found.

As mentioned above, the linear appearance of the cluster is also
apparent in the X-ray images, and, as seen in the image
of Fig.~\ref{xray}, this impression is given mainly by the presence of
the X-ray sources A, B, \& C in a line. These sources were also
present in past X-ray images (e.g., Nikogossyan et al. 1999).

Figure~\ref{xray} comes from the \xmm\ data, and it shows 
counts in the (0.5-5.0)\,keV energy range. \cl\ was observed by the
\xmm\ satellite for a total of $\sim$22~ksec on 2002 December 23--24
(revolution=557). During the observation the EPIC instruments were
operating in the PrimeFullWindow (for MOS1 and MOS2), and
PrimeFullWindowExtended (PN), and the thin filter was used for all
imaging detectors. We obtained the raw \xmm\ data from the data
archive, and processed them with SAS v6.5.0. {\sc emchain} and
{\sc epchain} were used to obtain the calibrated event lists for the
MOS and PN instruments respectively. The calibrated events were
filtered for {\it flag}s, using the \xmm\  {\it flag}s {\sc $\#$XMMEA\_EM}
and {\sc $\#$XMMEA\_EP} for the two MOS and the PN detectors
respectively. Restrictions on the {\it pattern} were also applied: we
kept only events with {\it pattern}$<$12 for the MOS cameras, and $<$4
for the PN. We also cleaned the event lists for periods of high
background, by applying a 3$\sigma$ clipping to the (10-15)\,keV lightcurves of each detector.  This cleaning process reduced the exposure times by
$\sim$2\,ksec. For all the spectral analysis, we used the clean and
filtered event lists, and we generated responses and auxiliary files
with {\sc rmfgen-1.53.5} and {\sc arfgen-1.66.4} respectively.
Subsequent spectral analysis was carried out in {\sc xspec}.

Recently, Mahdavi et al (2005) argued that the source C does not
belong to \cl\, but rather is a background cluster of galaxies at $z
\simeq 0.15$. Its central galaxy is at $\alpha = 01^{h}25^{m}35\fs9$
$\delta = -01\degr25\arcmin45\arcsec$, and appears much fainter than
the other cluster galaxies. The identification of source C with a
background X-ray source could make the linearity of the cluster a
questionable property. We used the processed \xmm\ data to derive the
X-ray properties of this source, in order to investigate further the
possibility of it being a background cluster. To obtain its spectrum,
we accumulated photons in a circular aperture centred on the peak of
its X-ray emission and extending out to 1\,arcmin. We fitted the
(0.5-5.0)\,keV spectrum with an absorbed {\it mekal} model. The
absorbing column ($N_{\rm H}$) was fixed to the Galactic column for
the direction of \cl\ ($N_{\rm H,G}=3.78 \times 10^{20} \ {\rm
cm^{-2}}$). The plasma temperature ($kT$), abundances ($Z$), and
normalization were left free to vary during the fitting procedure.
Additionally, the redshift was fixed to $z=0.15$, as indicated by the
work of Mahdavi et al. (2005). We obtained a temperature of $kT =
1.64^{+0.14}_{-0.18}$\,keV and a metal abundance of
$Z=0.48^{+0.27}_{-0.18} \, {\rm Z_{\odot}}$ for
$\chi^{2}/dof$=113/118. Adding a power-law component does not change
the best fitting model. The bolometric X-ray luminosity implied by the
best-fitting model for a source redshift of 0.15 is $L_x \simeq 0.8
\times 10^{43} \ {\rm erg \ s^{-1}}$, which would make it consistent
with observed cluster temperature-luminosity relations (Xue \& Wu
2000, Osmond \& Ponman 2004). If instead we fix the redshift to that
of \cl\ we find a temperature of $kT = 1.06^{+0.06}_{-0.06}$\,keV and
a metal abundance of $Z=0.13^{+0.05}_{-0.04} \, {\rm Z_{\odot}}$, and
obtain a marginally poorer fit ($\chi^{2}/dof$=127/118). We cannot
therefore rule out the possibility that source C is at the redshift of
A194 based purely on the X-ray data, but they are also certainly
consistent with it being a background cluster, and given the results
of Mahdavi et al. we exclude it from the discussion that follows.

\begin{figure}
\begin{center} 
\leavevmode 
\epsfxsize 1.0\hsize
\epsffile{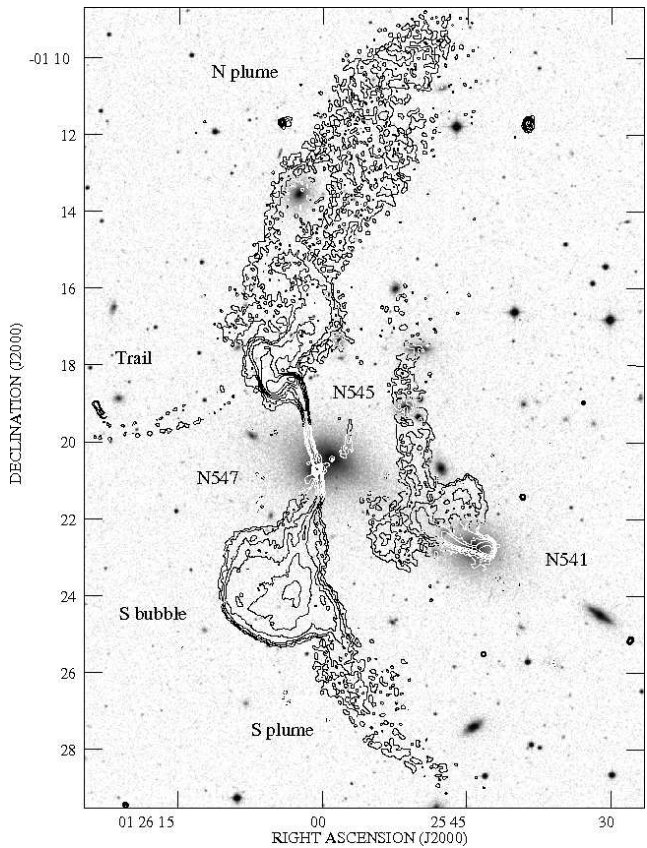}
\caption{A $5.6 \times 5.2$ arcsec resolution 1.5-GHz VLA map
  superposed on the DSS2 $R$-band image of the field of 3C\,40. Grey
  levels are logarithmic. Contours are at $300 \times (1,2,4\dots)$
  $\mu$Jy beam$^{-1}$. The three most massive galaxies of the cluster
  are labelled: note that each is a radio source, though the source
  associated with NGC 545 is very weak. Features of the NGC 547 WAT
  are also labelled.}
\label{OptRad}
\end{center} 
\end{figure}

Source A (see Fig.~\ref{xray}) is the dumb-bell galaxy NGC\,545/7
(Fig.~\ref{OptRad}). The member galaxies, NGC 545 and NGC 547 are of
similar magnitudes, have recession velocities $(5338 \pm 7)\ {\rm km \
s^{-2}}$ and $(5468 \pm 6)\ {\rm km \ s^{-1}}$ respectively (NED), and
are separated spatially by a projected distance of
$\sim$30\,arcsec=10\,kpc. Although their close separation and similar
velocities make it very likely that they are bound and moving within a
common gravitational potential, past observations have not revealed
strong evidence of interactions in the form of tidal tails and/or
bridges between the two (Fasano, Falomo, \& Scarpa 1996). The host of
the WAT radio galaxy is the SE member of the pair, the galaxy NGC 547.

Source B is identified with the galaxy NGC\,541, which hosts the NAT
radio galaxy PKS 0123-016A. This radio galaxy may have been
responsible for triggering star formation in the nearby `Minkowski's
object' (Brodie et al. 1985, van Breugel et al. 1985, Croft et al.
2006). Its recessional velocity is very close to that of the dumb-bell
system [$(5422 \pm 6)\ {\rm km \ s^{-1}}$ (NED)], and the galaxies are
separated by $\sim$4.5 \ arcmin = 97.5 \ kpc in projection.

Given the small differences of their velocities and separation it
might appear likely that the dumb-bell galaxy and NGC\,541 constitute
a bound system, possibly orbiting around a common centre of gravity.
Optical observations reveal an optical `bridge' that connects
NGC\,545/547 and NGC\,541 (see fig.\,4 in Croft et al. 2006), which
could be a sign of past and/or on-going interactions between them. The
issue of the interactions and dynamics of these galaxies is discussed
further in Section\,4.

For the purpose of any subsequent spatial analysis, we firstly created
exposure-corrected images for each EPIC instrument in the
(0.5-5.0)~keV energy range. The exposure correction was performed as
in Sakelliou \& Ponman (2004). These images from each \xmm\ camera
were fitted in {\sc sherpa} with a 2-dimensional $\beta$-model. We also
added a 2-dimensional constant to represent the background, which was
not subtracted from the images. The fit was restricted to the inner
10~arcmin of the image, and all bright point sources were subtracted.
We also subtracted the sources A, B, \& C (Fig.\,1) by masking out
circular regions with radius of 1\,arcmin around them. The images from
the three \xmm\ cameras were fitted simultaneously. The core radius
($r_{\rm c}$), the $\beta$-parameter and the location of the X-ray
centre were linked and left free, so that their values are determined
by the fitting procedure. The normalizations of the $\beta$-models
were free to vary independently for each instrument. This procedure
resulted in the following best fitting parameters: $r_{\rm c}$
=5.14$_{-0.0924}^{+0.133}$~arcmin=111.40$_{-0.195}^{+0.199}$~kpc, and
$\beta$ =1.11$\pm$0.19. The X-ray centre is found to lie at $\alpha_{cen} =
01^{h}25^{m}50\fs70$ $\delta_{cen} = -01\degr22\arcmin10\farcs80$,
which does not coincide with any of the bright cluster galaxies.

Source spectra were extracted in a circular region centred near the
cluster centre found in the 2-dimensional analysis, and extending out
to 5~arcmin. The background was taken from an annular region adjacent
to the source region, between 5 and 8~arcmin from the cluster centre.
Point sources, the emission around the galaxies A and B, and source C
(see Fig.\,1) were excluded. Spectra from the three \xmm\ instruments
were fitted simultaneously with a {\it mekal} model modified by the
absorbing column ($N_{\rm H}$). During the fitting procedure, the
$N_{\rm H}$ and metallicity ($Z$) were held fixed to the Galactic
value ($N_{\rm H}=N_{\rm H,G}=3.78 \times 10^{20} \ {\rm cm^{-2}}$)
and $Z=0.3Z_{\sun}$ respectively. The temperature of the plasma
($kT$), and the normalization were free to vary. We found a
temperature of $kT=2.87^{+0.33}_{-0.29}$
($\chi^{2}/d.o.f.=822.3/695$). The temperature we derived is in good
agreement with the {\it ASCA} results (Fukazawa et al. 1998), but
larger than the value of (1.36$\pm$0.04)~keV found by Mahdavi et al.
(2005). In order to compare with the $L_{\rm x} -T$ relation of
similar clusters we extrapolated the luminosity of the above model out
to $R_{500}$. We find a bolometric luminosity of $L_{\rm x}(R_{500})
\sim 2 \times 10^{43} \ {\rm erg \, s^{-1}}$, which makes our value
for the cluster temperature consistent with the $L_{\rm x}-T$ relation
presented by Osmond \& Ponman (2004).

\section{The radio source 3C\,40}

\subsection{Radio observations}

Abell 194 hosts the bright radio source 3C\,40 (PKS 0123$-$016). In
fact 3C\,40 consists of two radio galaxies, both of which contribute
to the catalogued 178-MHz flux density of 3C\,40 (e.g. Maltby,
Matthews \& Moffet 1963). The brighter source, which is usually known
as 3C\,40 but should properly be referred to as 3C\,40B, (0123$-$016B)
is a WAT (see Hardcastle \& Sakelliou 2004 for a definition) hosted by
NGC 547, while the better-known 3C\,40A (0123$-$016A) is a NAT hosted
by NGC 541, and was studied in the radio by van Breugel et al. (1985)
and Brodie et al. (1985). Detailed radio studies of the whole cluster
have not been carried out to date, but images of the field have been
presented by Brodie et al. (1985) and O'Dea \& Owen (1985: VLA
snapshot data) and by Nikogyossian et al. (1999: NVSS data). Our
attention was originally drawn to the cluster, part of the sample of
Jetha, Hardcastle \& Sakelliou (2006), because the WAT 3C\,40B is
clearly not at the cluster centre as defined by the diffuse X-ray
emission, unlike almost all other WATs we have studied. Location at
the centre of a cluster has been thought to be a key feature
determining WAT morphology (e.g. Hardcastle 1998) and so it is
important to understand why 3C\,40 is capable of supporting a
non-cluster-centre WAT.

To investigate the relationship between the complex structure of the
radio galaxies and the X-ray emitting ICM we retrieved deep VLA
imaging data at 330 MHz and 1.5 GHz from the VLA archive. Details of
the VLA observations are given in Table \ref{vla}: the 1.5-GHz
observations are the same as those presented by Jetha et al.\ (2006).
Observations at higher frequencies exist, but because of the large
angular scale of the source (24 arcmin) they cannot map the
large-scale structure adequately. We used archival VLA data at 4.9 GHz
(Table \ref{vla}) solely in order to make accurate measurements of the
bending of the jets. In addition, we observed 3C\,40 with the GMRT at
240 MHz and 610 MHz (in dual-band mode) for 8 hours on 2006 Jul 21.
The data were reduced in the standard manner within {\sc aips}, with
particular care being taken to flag data affected by radio-frequency
interference (RFI) at low frequencies. At all frequencies below 1.5
GHz the data were taken in spectral line mode, and (although spectral
channels were averaged before imaging) multiple channels were retained
throughout the analysis: imaging was carried out using the task {\it
imagr} with evenly-spaced grids distributed over the primary beam to
allow the removal of background point sources, and one or two
iterations of phase self-calibration were used before the final maps
were produced. The full-resolution 1.5-GHz VLA map has a resolution of
$5.6 \times 5.2$ arcsec (resolutions here are the FWHM of the
elliptical, or, where only one number is quoted, circular restoring
Gaussians) and an off-source noise of 60 $\mu$Jy beam$^{-1}$: the
330-MHz map has a resolution of $22 \times 19$ arcsec and an
off-source noise level of 1.7 mJy beam$^{-1}$. In general, the
off-source noise in the GMRT maps is good (at 240 MHz, the nominal
level is 0.8 mJy per $19 \times 15$ arcsec beam) but the image
fidelity is not as good as in the VLA images, with more `lumpiness' in
the maps: we attribute this partly to residual RFI and partly to the
poorer $uv$ plane coverage of the GMRT at the declination of 3C\,40.
The GMRT data, however, are adequate to make low-resolution maps for
spectral index comparisons.

\begin{figure*}
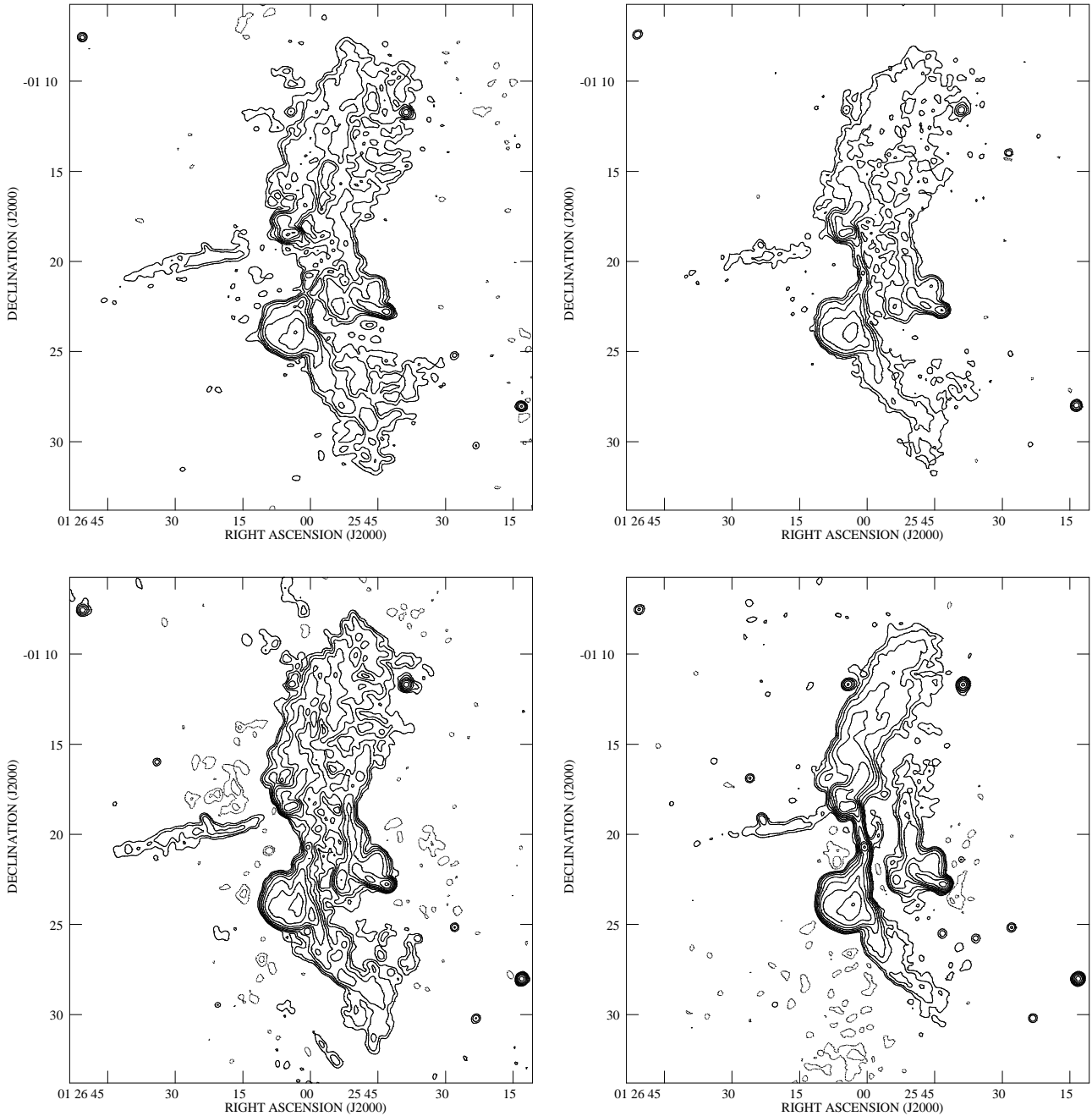

\begin{center} 
\leavevmode 
\epsfxsize 0.49\hsize
\epsffile{3C40-20-240.PS}
\epsfxsize 0.49\hsize
\epsffile{3C40-20-330.PS}
\end{center}
\begin{center}
\leavevmode 
\epsfxsize 0.49\hsize
\epsffile{3C40-20-610.PS}
\epsfxsize 0.49\hsize
\epsffile{3C40-20-1540.PS}
\caption{3C\,40 at 20-arcsec resolution. Top left: 240-MHz GMRT data.
 Top right: 330-MHz VLA data. Bottom left: 610-MHz GMRT data. Bottom
 right: 1.5-GHz VLA data. Contours are at $(-2,-1,1,2,4\dots)$ times
 the $5\sigma$ off-source noise level on each map: these $5\sigma$
 levels are respectively 5.0, 5.8, 1.1 and 0.59 mJy beam$^{-1}$.}
\label{vla-large}
\end{center} 
\end{figure*}

Fig. \ref{OptRad} shows the full-resolution 1.5-GHz VLA image
superposed on the DSS-2 $R$-band optical image. This image illustrates
several interesting features of the radio source. Note in particular
the weak radio source associated with NGC 545, with a short jet that
crosses the much more prominent jet of NGC 547, and the double-peaked
trail that extends E from the N plume of the NGC 547 WAT (discussed in
more detail in Section \ref{trail}). In Fig.\
\ref{vla-large} we show maps at all four frequencies, made with a
common resolution of 20 arcsec. These maps are made from the images
used for spectral index analysis. It can be seen that all images sample
the same structure (though weak negative contours at 610 and 1540 MHz
suggest that there may be some slight undersampling on the largest
scales) but that the VLA data give generally smoother and more
physically plausible maps.

\begin{table}
\caption{VLA archival observations of 3C\,40 used in this paper}
\label{vla}
\begin{tabular}{lllll}
\hline
Frequency&Obs. date&VLA obs.&Config.&Time on\\
(GHz)&&ID&&source\\
&&&&(minutes)\\
\hline
4.9&1984 Jan 08&AV102&B&156\\
4.9&1984 Jun 02&AV102&C&212\\
1.5&1984 Jan 08&AV102&B&335\\
1.5&1984 Jun 02&AV102&C&206\\
1.5&1985 Jul 31&AV112&D&81\\
0.33&1994 Aug 16&AE97&B&316\\
0.33&1994 Nov 20&AE97&C&40\\
\hline
\end{tabular}
\end{table}

\subsection{The WAT \& NAT hosts}

The jet and plume morphologies of the radio galaxies suggest that
their host galaxies are moving relative to the ICM. Such motions
relative to the ICM could result in the stripping of their
interstellar media (ISMs). However, the images make clear that there
is small-scale X-ray emission associated with both NGC 547 and NGC
541. In order to investigate if they retain massive ISMs, we obtained
the X-ray spectra of the WAT and NAT hosts within circular regions
with radii of 40\,kpc = 110\,arcsec. For each galaxy, the background
was taken in an annulus adjacent to the source region extending from
110 to 130 \,arcsec, and we masked out background point sources and
(in the case of NGC 547) the X-ray emission associated with the
companion galaxy. The spectra were modeled in {\sc xspec} by {\it
mekal+power law} models absorbed by the Galactic column, where the
power-law component takes account of the expected unabsorbed
jet-related nuclear emission from the FRI radio galaxies. The
best-fitting temperatures were well determined and significantly lower
than the temperature for the cluster as a whole ($0.61 \pm 0.03$) keV
for NGC 547 and ($0.95 \pm 0.07$) keV for NGC 541. From the
normalization of the {\it mekal} models (with fixed abundance of 0.3
solar) and assuming that the ISM is spherically symmetric, we find
that $M_{\rm ISM, WAT} = 4.50^{+0.12}_{-0.20} \times 10^{9} \
M_{\sun}$, and $M_{\rm ISM, NAT} = 3.98^{+0.36}_{-0.39} \times 10^{9}
\ M_{\sun}$, values that are comparable to the gas masses of other
massive elliptical galaxies (e.g., Canizares, Fabbiano \& Trinchieri
1987). In addition, we carried out fits of models consisting of a
central point source and isothermal $\beta$ model (both convolved with
the \xmm\ PSF) to the radial profiles of the two galaxies within
the 110-arcsec regions, with the same masking of unrelated sources.
These give plausible small core radii and $\beta$ values (both
galaxies have best-fitting $\beta \approx 0.5$) and the masses derived
by integrating the $\beta$ models are very consistent with those
derived from the {\it mekal} normalization.

The masses of the ISMs we found above, indicate that both the WAT and NAT hosts still retain large amounts of their ISM. It seems, that ram pressure has not been efficient in stripping them from their hot haloes.


\section{Discussion}

\subsection{Spectral ageing and the `trail'}
\label{trail}

Both of the powerful radio sources in \cl\ show clear radio spectral
steepening along their length, as is generally observed in FRI radio
sources. The spectral steepening is normally attributed to the
progressive loss of high-energy electrons via synchrotron and
inverse-Compton losses. However, using the radio spectra to determine
the time since the electrons at various points in the plume had a flat
spectrum, the `spectral age', is well known to be fraught with
difficulties: the method assumes that there is no diffuse particle
acceleration throughout the plume, it requires the assumption of a
model for electron diffusion (e.g. pitch angle diffusion, Jaffe \&
Perola 1973), it requires that we assume a single electron energy spectrum
(whereas there is some evidence for multi-component structures in the
plumes of some FRI sources: Katz-Stone \& Rudnick 1997, Hardcastle
1999), and it requires that we know the magnetic field strength and,
in general, the {\it history} of the magnetic field strength as seen
by the electrons. In regions where the field is sufficiently weak that
the magnetic field energy density is much less than that of the CMB,
$B^2/2\mu_0 \ll 8\pi^5(kT_{\rm CMB})^4/15(hc)^3$, though, losses to
inverse-Compton scattering can be presumed to dominate and the last of
these requirements is removed.

In spite of these problems it is worthwhile to estimate the spectral
ages of parts of the radio source in 3C\,40. While the numerical
values of age that we derive will be model-dependent, they
nevertheless provide a useful comparison to other ages of the system
derived from dynamical constraints. In addition, we can use the ages
to help us understand the structure of the radio source.

We measured flux densities from relatively large regions of the radio
sources, and fitted Jaffe \& Perola (1973) aged synchrotron spectra to
them, assuming an electron injection index $p$ [where $N(E) \propto
E^{-p}$] of 2.1 (as recently argued by Young et al.\ 2005). We assumed
that the magnetic field had the equipartition value, modelling the
emission region as a cylinder in the plane of the sky. In the outer
parts of the plumes of the WAT, where the spectrum is steepest, the
equipartition magnetic field energy density is well below the CMB
photon energy density, so that inverse-Compton losses dominate. Thus, 
the age that we derive by
   this method for the outer part of the WAT plumes, $1.5 \times
10^8$ years, is only weakly dependent on the magnetic field used
(going roughly as $B^{1/2}$) and so is very little influenced by
moderate departures from equipartition. This value of the age is
essentially an upper limit, since the ageing $B$-field may have been
higher (if equipartition is maintained, the $B$-field certainly was
higher in the brighter, higher-pressure regions of the plume) and
since we take no account of the effects of adiabatic expansion of the
plumes, which mimics ageing. We can thus say that the time since the
last particle acceleration event is $< 1.5 \times 10^8$ years: if
particle acceleration occurred predominantly at the plume base, then
the material at the ends of the plume has taken $< 1.5 \times 10^8$
years to travel there, implying travel speeds of $>1400$ km s$^{-1}$
(where the lower limit on speed takes into account both the limit on
age and the effects of projection).

Spectral ageing analysis also sheds some light on the origin of the
`trail' seen to the E of the N plume of the WAT. The spectral index
of this material is much steeper than that of the region of the WAT
immediately adjacent to it: this implies that it is not directly
related to the WAT plume. The spectral index of the trail steepens
with increasing distance eastward, and the spectral age of the
steepest-spectrum material (which should be interpreted as for the
WAT) is $1.4 \times 10^8$ years. The difference between the spectra of
the trail and the adjacent WAT material, together with the
double-peaked radio structure of the trail, strongly suggests that the
trail is not related to the WAT. Instead it must originate in one of
the other two radio galaxies in the system. The spectral index of the
NAT material on the opposite side of the WAT to the trail is
comparable to -- in fact slightly steeper than -- that of the material
at the E end of the trail, and there is some evidence in the radio
maps that the plume extending to the N of the NAT has the same
double-peaked structure as the trail. In this picture the material
produced by the NAT has to make a sharp (90$^\circ$) bend in
projection to form the trail. The alternative is that the
trail was produced by the weak radio source that we associate with NGC
545, with the connection between the host galaxy and the trail being
obscured by the bright intervening WAT emission. The lack of any
obvious connection in the radio between NGC 545 and the trail argues
against this idea, however.

\subsection{Dynamics of the cluster centre}

All the data presented in the previous sections show that the core
region of \cl\ is occupied by large galaxies. These galaxies are
currently in close separation, they host bent radio sources, and live
in an elongated X-ray envelope. Taking into account these
observational facts, and past beliefs that the existence of tailed
radio galaxies in clusters is closely connected to the disturbances
induced by recent cluster mergers, as more tailed sources are found in
clusters that have suffered a recent merger event (e.g., Bliton et al.
1998), one could naturally think that \cl\ has been recently assembled
by the accretion of smaller mass units. In such a formation scenario
the radio galaxies could be newcomers to the cluster, having fallen
into it from its outskirts; now they should be crossing its central
region at high speeds. If this picture is correct, their radio jets
should be bent by the resulting high ram pressure. It is also accepted
that they should not retain large amounts of ISMs, as they should have
been stripped during their in-fall towards the cluster centre.
Luckily, we can test all the above using the information from the
X-ray and radio data.

The shape of the jets, for example, puts constraints on the galaxy
velocities, if we assume that the jets of both the NAT and the WAT are
bent by ram pressure, and apply the usual equation: $\rho_{\rm j}
v_{\rm j}^2/R = \rho_{\rm ICM} v_{\rm gal}^{2}/h$, where $h$ is the
scale height of the jet and $R$ is its radius of curvature. The
structure of the NAT clearly suggests bending by bulk motion. The WAT
jets, though bent, are less obviously symmetrically swept backwards,
but we estimate the possible effects of bending from the prominent
bend in the S jet between 15\,arcsec and 1\,arcmin from the nucleus.
We measured the scale heights ($h_{\rm NAT}$, $h_{\rm WAT}$) and radii
of curvature ($R_{\rm NAT}$, $R_{\rm WAT}$) from high-resolution radio
maps, obtaining $h_{\rm NAT}$ = 4\,arcsec, $h_{\rm WAT}$ = 1\,arcsec,
$R_{\rm NAT}$ = 37\,arcsec, $R_{\rm WAT}$ = 65\,arcsec. For the
density of the ICM near the location of both radio galaxies we used
the $\beta$ model fits described above and the methods of Birkinshaw
\& Worrall (1993) to determine a value of $n_{\rm p, ICM} = 5 \times
10^{-4} \ {\rm cm^{-3}}$: the dominant uncertainty in this quantity is
the systematic uncertainty from the assumed abundance, but for a
plausible abundance range at our best-fitting cluster temperature the
fractional uncertainty is of the order 20\%. The most uncertain
quantities in the above equation are the densities and velocities of
the jets. Lower limits on the densities may be estimated from the
minimum energy density, which is $7 \times 10^{-13}$ J m$^{-3}$ for
the NAT and $3 \times 10^{-12}$ J m$^{-3}$ for the WAT. Since there is
no sign of beaming in the NAT jet, we take $v_{\rm j, NAT} = 0.1c$, a
reasonable value for the jet speed in FRI jets after deceleration
(Laing \& Bridle 2002) while for the WAT we take $v_{\rm j, WAT} =
0.5c$, the best-fitting value for WAT jet speeds derived by Jetha et
al. (2006), and assume the jet is in the plane of the sky. This gives
limits of $v_{\rm gal, NAT} > 30$ km s$^{-1}$ and $v_{\rm gal, WAT} >
110$ km s$^{-1}$. These are lower limits because we do not know the
true jet densities. For NATs, like other normal twin-jet FRIs, we
expect deceleration to have taken place by entrainment, so that the
true density could be substantially above the minimum value. The
models of Laing \& Bridle (2002), applied to the jet in 3C\,31,
require a jet density contrast ($\rho_{\rm jet}/\rho_{\rm ICM}$) of
$10^{-4}$ on 10-kpc scales, which in NGC 541 would correspond to
$v_{\rm gal, NAT} = 140$ km s$^{-1}$. If the speed of NGC 541 with
respect to the ICM were much higher than this, the jet would have to
be substantially heavier with respect to the ICM than that in 3C\,31,
and/or substantially faster than seems likely, in order to avoid being
more strongly bent by ram pressure.

The above calculations suggest that the speeds of the WAT and NAT
hosts should be low compared to the trans- or super-sonic velocities
they would have had if they have fallen recently into the cluster (see
for example fig.\,1 in Acreman et al. 2003). As was mentioned in
Section\,2 past measurements of their recessional velocities also
found a low velocity separation along the line of sight. 

Such low velocities are not generally adequate to strip the galaxies from their ISMs.   
Acreman et al. (2003) simulated the stripping process during the
infall of elliptical galaxies towards the centre of a cluster. As their simulated cluster had a temperature of
kT=2.7\,keV, it is very comparable to \cl.
As can be seen in their fig.\,1 the galaxy falls from the outskirts of the cluster towards the cluster centre. Its velocity increases and reaches supersonic velocities ($\ge1000\,{\rm km \ s^{-2}}$) only around the cluster core region. In all cases they studied, the galaxies do not retain
more than $\sim 4 \times 10^{9} \ M_{\sun}$ of their ISM; the gas mass
that remains within these galaxies can be as low as $\sim 10^{8} \ M_{\sun}$.
Only the ISMs of very massive galaxies are not severely stripped:
galaxies with dark matter haloes as large as $M_{\rm halo}= 4 \times
10^{12} \ M_{\sun}$ do not suffer any substantial reduction of their
ISMs. Severe stripping occurs only when the galaxy is crossing the dense cluster core cluster region at high velocities. 
The gas masses we calculated in Section\,3.2 strongly suggest that that
the radio galaxies in \cl\ have not been stripped of their X-ray
halos, and therefore they cannot be currently falling towards the
cluster centre from its outskirts. Thus, the lack of stripping  also argues against a scenario in which  the core region of \cl\ is undergoing a major merger event.

We are therefore forced to the conclusion that the NAT and WAT hosts are moving at low speeds. The two radio galaxies are at 
projected distance of $\sim$50\,kpc from the
cluster centre. The gravitating cluster mass within such a radius is
$\sim 1 \times 10^{12}\, {\rm M_{\sun}}$ (Nikogossyan et al. 1999).
If they are bound their velocities should be $\leq 300\,{\rm km \
s^{-1}}$, which is within the allowed limits of the galaxy velocity
found above. Thus, we conclude that they are bound to, and probably
rotating around, the cluster centre. The expected orbital speed is
adequate to bend the jets while still retaining the observed amount of
hot gas within the galaxies.

It should also be noted that there is nothing unusual about the
distribution of the cluster gas. The X-ray data do not show any signs
of features expected from a recent cluster merger event, such as regions
of enhanced emission due to shock waves or small in-falling groups of
galaxies.

Thus it appears that these cluster galaxies are most likely orbiting
around the common centre of mass inside the cluster core. 
The tips of the radio galaxies are probably not passive radio trails
left behind in the cluster as the galaxies have moved to their current
positions. It would have taken 1.2 Gyr for the NAT, for example, to
create the North trail as a passive trail (assuming motion of 6
arcmin, or 130 kpc, at 100 km s$^{-1}$) and this is inconsistent with
the observed spectral ages (which, as discussed above, should probably
be taken as upper limits). Similarly, the initially attractive idea
that the bending of the plumes of the WAT (in the opposite sense to
the current bending of the inner jets) is a result of their having
being produced at a stage when the host galaxy was moving in the
opposite direction is inconsistent with the very long orbital period
of the system (the current period is $T\simeq1$\,Gyr), unless there is
substantial particle reacceleration in the plumes. In our picture, in
which the host galaxies are moving slowly and have remained close to
the cluster centre for much longer than the radio source lifetimes,
the bending of both WAT and NAT plumes must be a result of the
interaction between the plume flow (transport speeds must still be
substantial if the observed spectral ages are accurate, see above) and
its environment. As discussed by Hardcastle, Sakelliou \& Worrall
(2005) interactions between a fast plume and external material with
sub-sonic speeds can only have a significant effect if the plume is
very light compared to the external medium, and in this case buoyant
forces probably also play an important role.

\section{Summary and conclusions}

We have presented \xmm, VLA, and GMRT observations of the nearby cluster \cl.
Its core region is occupied by the WAT 0123$-$016B, and the NAT
0123$-$016A, which together comprise the radio source 3C\,40.

The X-ray data do not show any signs of features expected from recent
cluster merger activity, like regions of enhanced emission due to
shock waves or small in-falling groups of galaxies. The X-ray and
radio data argue that the jets and plumes of the radio sources do not
require high galaxy velocities to be bent into their current shapes.
Relatively low velocities of the order of a few hundred ${\rm km \
s^{-1}}$ are adequate to produce the bending. Additionally,
both galaxies retain a large fraction of their ISM, suggesting that
they are not severely stripped as it would have been the case if they
are newcomers to the cluster having fallen from its outskirts. The
WAT's host is not at the cluster centre as defined by the distribution
of the ICM.

It appears that \cl\ is not suffering a major merger event, that would involve large relative galaxy velocities. A likely
scenario is that the radio galaxies are rotating around the cluster
centre in the core region of \cl. Such a motion could explain the
observational facts like the bending of the jets/plumes of the radio
galaxies, the presence of the optical bridge, and the lack of
stripping of the galaxy haloes. The presence of the two massive X-ray
haloes and the lack of rapid motions of the host galaxies probably
explain how there can be two powerful radio galaxies so close to the
centre of the cluster and why 3C\,40B can have a classical WAT
morphology in spite of its offset from the cluster centre.


\section*{Acknowledgments}

We thank an anonymous referee for constructive comments that helped
us to improve the paper.
The Digitized Sky Survey (DSS), and the NASA/IPAC Extragalactic
Database (NED) have been used. The present work is based on
observations obtained with \xmm, an ESA science mission with
instruments and contributions directly funded by ESA Member States and
the USA (NASA). The National Radio Astronomy Observatory is a facility
of the National Science Foundation operated under cooperative
agreement by Associated Universities, Inc. We thank the staff of the
GMRT for their help with the observations with that telescope: GMRT is
run by the National Centre for Radio Astrophysics of the Tata
Institute of Fundamental Research, India. IS acknowledges the support
of the European Community under a Marie Curie Intra-European
Fellowship. MJH thanks the Royal Society for a research fellowship.


\end{document}